\documentstyle[12pt]{article}

\begin{document}
\title{Light Front Dynamics and The Binding Correction}
\author{S.V.Akulinichev \thanks{E-mail address: AKULINICHEV@INUCRES.MSK.SU
and SERGUEI@TAMCOMP.BITNET}\\
{ \it Institute for Nuclear Research, 60-th October Anniversary}\\
{ \it Prospect 7a, Moscow 117312, Russia}\\
and\\
{ \it Cyclotron Institute of Texas A\&M University,}\\
{ \it College Station, TX 77843-3366, USA} }
\date{March, 1993}
\maketitle
\begin{abstract}
It is shown that the nuclear binding correction in deep inelastic
lepton scattering
is essentially the same in light front and instant form representations.
Some contradicting papers are discussed.
\end{abstract}
\newpage

In the earlier works\cite{We1,We2} it has been shown that the bulk of the
nuclear EMC-effect in deep inelastic lepton scattering (DILS) can be explained
by the binding correction which leads to the x-rescaling of structure
functions.
This result has restricted the possible contribution of more exotic effects.
However, in some later papers it was claimed that the binding correction is
unimportant in nuclear DILS. For example, the authors of
Refs.\cite{BC}-\cite{JK
}
 conclude that the binding correction almost vanishes if the light front (LF)
formulation is used instead of the instant form (IF) one. This point of view
was shared by several authors. Here it will be shown that the binding
correction
in the LF representation is essentially the same as in
the IF representation.
The aim of this note is to clear up both the role
of binding in the nuclear EMC effect and the connection between the LF and
IF descriptions of DILS.

In the LF representation,  the operator $\underline{P}\equiv
(P_{+},P_{1},P_{2})
$
belongs to the stability group that maps the quantization surface $x^{-}=0$
onto itself \cite{Dir,LS}. This means that for any physical system (not
necessarily free and/or localizable) the three-momentum $\underline{P}$
is among the observables: it is well defined, strictly conserved and
transformed according to the Lorentz transformations.
The momentum $\underline{P}$ in the LF representation is an  analog
of the usual three-momentum $\vec{P}$ in the IF representation. For an
interacti
ng
particle, the momentum $\underline{P}$ is the parameter on which the
interactions of this particle depend. Therefore
the observed momenta $\underline{P}$ of interacting particles and
their interactions are interconnected.

For any particle (with non-zero mass), the "Hamiltonian" $P_{-}$
can be rewritten by introducing the mass operator $M$,
\begin{equation}
P_{-} = (M^{2} + P^{2}_{1} + P^{2}_{2}) / 2 P_{+}.
\end{equation}
For free particles, the free mass $m$ is the sharp value of the mass operator
and in this case $P_{-}$ is the remaining component of the four-momentum.
However, for interacting particles, the four-momentum is not defined \cite{LS}
and only the three-momentum $\underline{P}$ is a defined physical value.
Therefore it is irrelevant to discuss whether or not the interacting particles
are on-mass-shell in the LF representation. One should not be
confused by the fact that in the derivation of the Weinberg equation
\cite{Wei},

that was often adopted to describe the LF dynamics, the
on-mass-shell energies of intermediate particles were used in the energy
denominators. It was shown\cite{CM} that the Weinberg equation is just
another form to represent the Feynman diagrams in terms of LF variables.
The fact that any Feynman diagram can be
rewritten as the sum of old-fashioned perturbation theory diagrams
does not mean  that interacting particles are on-mass-shell in the
time-ordered perturbation theory.
Hopefully, as  it is demonstrated here, the full four-momenta
of interacting nucleons are not needed to reveal the nuclear
binding effect.

In incoherent impulse approximation for the inclusive  scattering, the nucleus
structure function in the LF representation can be written in the
convolution form \cite{BC,Oel}
\begin{equation}
F^{A}_{2}(x) = \int dz\: f(z) \: F^{N}_{2}(\frac{x}{z}),\:\:\:\:\:
f(z)=\int d^{3}\underline{p}\: \delta (z - \frac{p_{+}}{m_{N}}) \rho
(\underline
{p}),
\end{equation}
where the nucleon momentum distribution function $\rho (\underline{p})$
is given by (see, e.g., Ref.\cite{Oel})
\begin{equation}
\rho(\underline{p})=\sum_{n, \underline{P}^{A-1}} \frac{\langle \underline{P}
^{A}| a^{+}(\underline{p})|n,\underline{P}^{A-1}\rangle \langle n,
\underline{P}^{A-1}| a(\underline{p})|\underline{P}^{A}\rangle}{A \langle
\underline{P}^{A}|\underline{P}^{A}\rangle}.
\end{equation}
By the sum in this equation
the integration over continuum states and momenta is understood.
The normalization condition
\begin{equation}
\int d^{3}\underline{p}\: \rho(\underline{p}) = 1
\end{equation}
follows from the definition. The structure function (2) can be directly used
to calculate the ratio $F_{2}^{A}(x) / F^{N}_{2}(x)$ that is determined in the
experiments.

The explicit form (3) of the momentum distribution function reminds us that
the inclusive cross section is the sum of squared amplitudes with
residual (A-1)-nuclei in the final state. This form  allows
to make a physical intepretation of the momenta:
$\underline{P}^{A}, \underline{P}^{A-1}$, and $\underline{p}$
are the LF momenta of initial nucleus, residual nucleus and struck nucleon,
respectively (as usual, the final state interactions are neglected).
Since the momentum $\underline{P}$ is strictly conserved , we have
\begin{equation}
p_{+} = P^{A}_{+} - P^{A-1}_{+}.
\end{equation}
We stress that within the adopted physical picture the LF state vectors
$|\underline{P}^{A}\rangle$ and $|n,\underline{P}^{A-1}\rangle$
describe the free localizable systems.
The LF four-momenta of these systems are well defined
and trivially related to the usual four-momenta. For example,
in the target nucleus rest frame
\begin{equation}
P^{A}_{+} = P^{A}_{0} = A(m_{N} - E_{B})
\end{equation}
where $E_{B}$ is the (positive) binding energy per nucleon. The states
$|n, \underline{P}^{A-1}\rangle$ are not observed in inclusive reactions.
By averaging over the velocity direction of the residual nucleus in the
target rest frame we obtain
\begin{equation}
\langle P_{+}^{A-1}\rangle \approx \langle P_{0}^{A-1} \rangle =
(A-1)(m_{N}-E'_{B}) + \langle E_{n}\rangle
+\langle E_{kin}\rangle,
\end{equation}
where $\langle E_{n}\rangle$ and $\langle E_{kin} \rangle$ are the average
excitation and kinetic energies of the (A-1) nucleus. For heavy nuclei
$E'_{B}\approx E_{B}$ and for the deuteron $E'_{B} =0$.

As it follows from (2), the nuclear structure function is determined by
$p_{+}$ of a struck nucleon. To estimate the nuclear binding effect,
it is convenient to use the expansion \cite{We2}
\begin{equation}
F^{A}_{2}(x) = F^{N}_{2}(\frac{x}{\langle z \rangle}) + \frac{1}{2}
(\langle z^{2} \rangle - \langle z \rangle^{2})\frac{\delta^{2}}
{\delta\langle z \rangle^{2}} F^{N}_{2}(\frac{x}{\langle z \rangle})+\ldots,
\end{equation}
where
\begin{equation}
\langle z \rangle = \int dz\:\:z\:f(z) = \frac{\langle p_{+} \rangle}{m_{N}}.
\end{equation}
For the qualitative estimates,
at $x \le 0.5$ it is sufficient to keep only the first term of this
expansion. From (5)-(9) we obtain
\begin{equation}
\langle z \rangle = 1 - (E_{B} +\Delta E_{B} + \langle E_{n} \rangle +
\langle E_{kin} \rangle ) / m_{N},
\end{equation}
where $\Delta E_{B} = (A-1)(E_{B} - E'_{B})$. The positive term in brackets
in the above equation can be recognized as the nucleon separation energy
$E_{sep}$.
For heavy nuclei
\begin{equation}
E_{sep}\approx 40 MeV,\:\:\langle z \rangle \approx 0.96
\end{equation}
and by (8)-(11) we recover the
earlier results for the nuclear binding correction \cite{We1,We2}.

It is possible to use the moment expansion introduced by Glazek
and Schaden \cite{GS}, as well,
\begin{equation}
F^{A}_{2}(x) = F^{N}_{2}(x) I_{1} + xF'^{N}_{2}(x)I_{2} + \ldots,
\end{equation}
where
\begin{equation}
I_{i} = \int dz\: f(z)\:(1-z)^{i-1}.
\end{equation}
{}From (2)-(4) and (10) we obtain $I_{1} = 1$ and
\begin{equation}
I_{2} = 1 - \langle z \rangle = \frac{E_{sep}}{m_{N}}.
\end{equation}
in agreement with the IF results.

In Refs.\cite{BC}-\cite{JK}, as well as in many other
papers on this subject, it was assumed
that in the LF representation there was an exact sum rule
\begin{equation}
\int d^{3}\underline{p}\: \rho (\underline{p})\: p_{+} = P^{A}_{+} / A.
\end{equation}
{}From (4), (13) and (15) it follows that
\begin{equation}
I'_{2} = 1 - \frac{m_{A}}{A m_{N}}.
\end{equation}
The moment $I'_{2}$ describes the binding effect in the approach of Refs.\cite
{BC}-\cite{JK} and for medium and heavy nuclei $I'_{2}$ is much less than
$I_{2}$ from (14). For the deuteron $I'_{2}\sim I_{2}$ since in this case
$\langle E_{n} \rangle \sim 0$ ( note that $I'_{2}$ from Ref.\cite{JK}
must be further corrected by the $m_{A} / A m_{N}$ factor as in the equation
(14) from Ref.\cite{BC}).

The sum rule  (15) can  be obtained  by keeping
only the nucleon part of the energy-momentum tensor \cite{SF}. But this part
of the energy-momentum tensor, without the meson  self-energy part,
is not sufficient to describe the bound system of nucleons.
By comparing (14) and (16) we can conclude that the binding correction
is roughly the measure of violation of the momentum sum rule (15) by meson
field
s.
The same situation can be found in one-nucleon case on the quark-gluon level:
the LF momentum sum rule is not fulfilled with only current quarks and
antiquarks, without the gluon contribution \cite{LS}.

I would like to thank the Cyclotron Institute of Texas A\&M University
for the warm hospitality.

\end{document}